\documentclass[useAMS,usenatbib]{mn2e}
\usepackage[OT4]{fontenc}
\usepackage{graphicx}

\title[Multiperiodic high-mass X-ray binary]{Rapid multiperiodic variability in a high-mass X-ray binary}
\author[Fabrycky]{Daniel Fabrycky\thanks{E-mail: dfab@astro.princeton.edu}\\
Princeton University Observatory, Princeton, NJ 08544-1001, USA}

\begin{document}

\date{Accepted ??. Received 2004 July 29}

\pagerange{\pageref{firstpage}--\pageref{lastpage}} \pubyear{2004}

\maketitle

\label{firstpage}

\begin{abstract}
Positions of High-Mass X-ray Binaries are often known precisely enough to unambiguously identify the optical component, and a number of those stars are monitored by the OGLE and MACHO collaborations.  The light curves of two such candidates are examined for evidence of Be star behaviour and for periodicity.  One of the stars exhibits two periods of 6.833 and 15.586 hours, much shorter and more stable than periods of Be/X-Ray Binaries that are attributed to the Be star's disk, but consistent with short-term Be variability attributed to pulsations.  The multiperiodicity is quantified with Fourier techniques and examined for phase stability; a combination of radial and non-radial pulsations is discussed.
\end{abstract}

\begin{keywords}
stars: oscillations -- stars: emission line, Be -- X-rays: binaries --  methods: data analysis
\end{keywords}

\section{Introduction}
Due to heavy coverage by recent X-Ray space missions, the number of known High-Mass X-ray binaries (HMXB) in the Small Magellanic Cloud (SMC) is rapidly expanding, and is now in the sixties \citep{b1}.  The more recent space missions, Chandra and XMM-Newton, are able to deliver positions accurate to a few arcseconds, opening the door to unambiguous optical identifications.  

Meanwhile, in the optical community, dense starfields are being monitored every few nights for variability with the goal of detecting massive compact objects via gravitational microlensing.  Many other areas of astronomy likewise benefit from well-sampled light curves in the optical and near-IR.  The MACHO project ran for seven years, until 1999, and has recently put its data online as a service to the community.  The OGLE project is ongoing, is currently in its third stage, and is expanding the area covered with each successive realization.

Careful analysis of a number of these optical counterparts to X-ray sources, which are frequently Be stars, is yielding interesting multiperiodic behaviour. For instance, \citet{b6} give four examples of stars which have an orbital period, judged by outbursts, of order one hundred days as well as quasi-periodic variations of order 10 days.  A region on a disk excited by the passing compact object may cause this latter variation as it orbits the star.

\citet{b2} identified two new X-ray pulsars with bright stars in the OGLE catalog of the SMC \citep{b3}.  I examine these stars for periodic variations, finding no result for the star associated with the 499.2 $s$ pulsar XMMU J005455.4-724512 (``star A'') but finding two periods that are both less than one day for the star associated with the 701.6 $s$ pulsar XMMU J005517.9-723853 (``star B'').  In section 2 I explain how the light curves were constructed.  A search for significant periods is covered by section 3.  Section 4 deals with the evolution of those periods over the time-scale of the observations.  I close with a discussion of the possible physical mechanisms causing this multiperiodicity in section 5.

\begin{table}
\begin{tabular}[b]{l|c|c|c}
 &XMM-Newton XMMU&OGLE SMC&MACHO\\
\hline
 Star A&J005455.4-724512&SC7 47103 &207.16254.16\\
 Star B&J005517.9-723853&SC7 129062&207.16313.35\\
\hline
\end{tabular}
\end{table}

\begin{figure}
\scalebox{.6}{\includegraphics{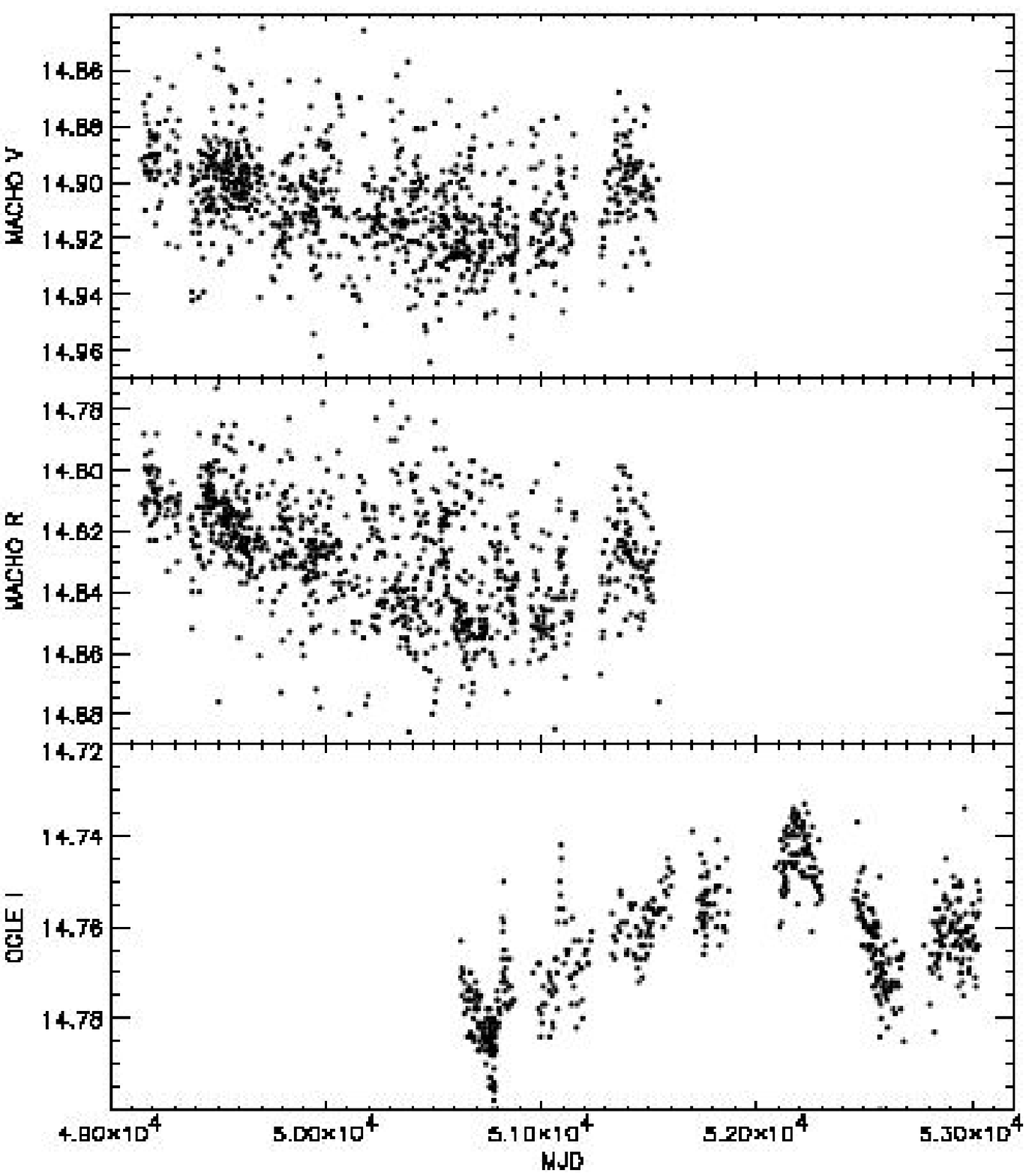}}
\caption{These are the {\it V} and {\it R} MACHO and {\it I} OGLE light curves for star A, which is associated with XMMU J005455.4-724512.  Errors are omitted for clarity.}
\scalebox{.6}{\includegraphics{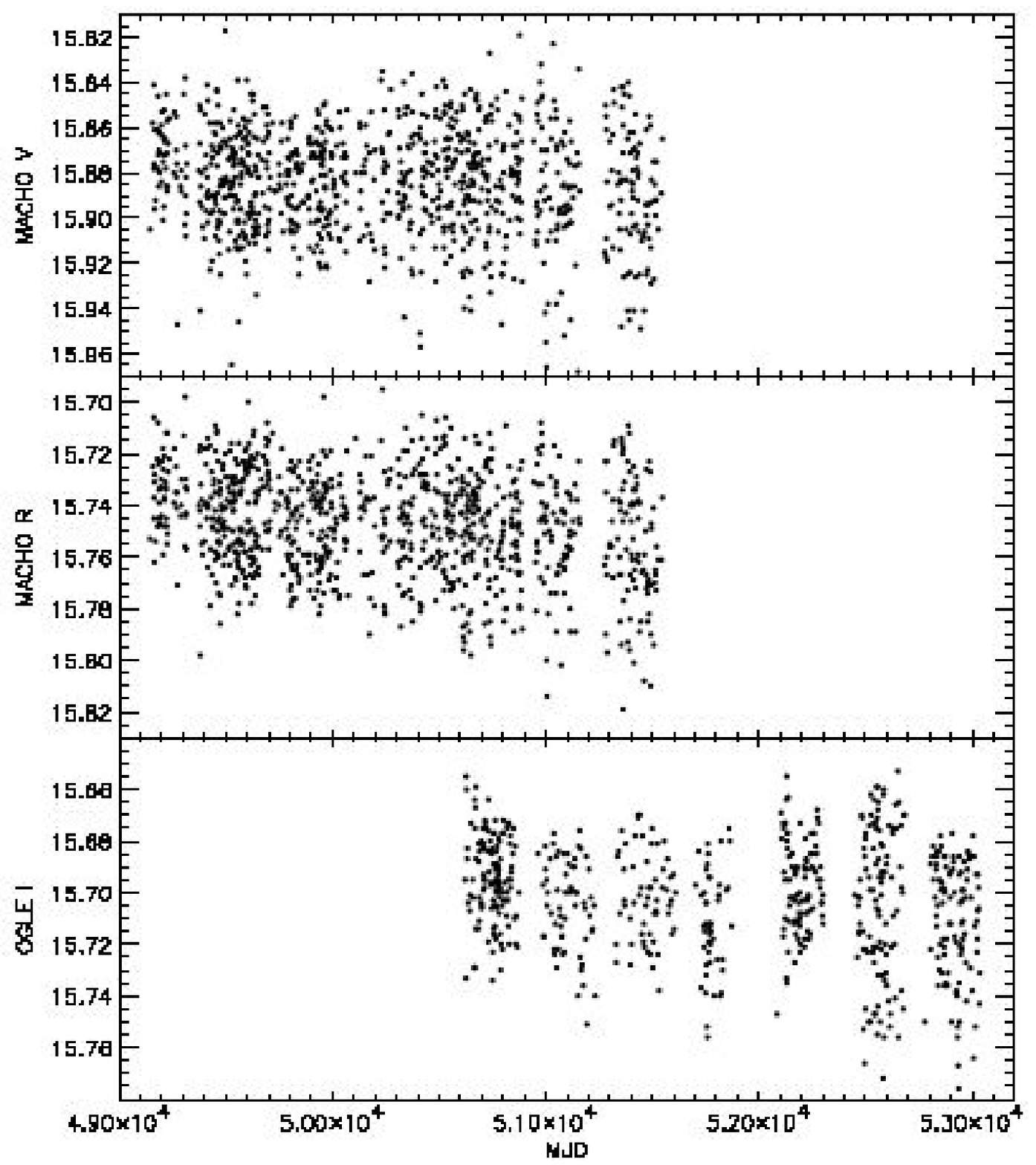}}
\caption{Same as above, but for star B, which is associated with XMMU J005517.9-723853.}
\end{figure}

\section{Light curves}
I obtained OGLE data for both stars and for nearby calibration stars from Andrzej Udalski.  This paper is the first time the OGLE light curves for these two objects have been published, since their amplitudes of variability are too small for membership in the catalog of SMC variable stars \citep{b4}.  They include both OGLE II and the ongoing OGLE III data, with the break at MJD 52000.  At the present time OGLE III is not carefully calibrated, so I selected stars within 10 arcseconds of the stars of interest that were brighter than magnitude 20 in the OGLE II catalog \citep{b3} to make a crude calibration.  I compared the difference in average magnitude of each star across the OGLE II/OGLE III break, then determined an offset to apply to the break by inverse variance weighting those individual offsets.  These total offsets have errors of 15 and 32 mmag for star A and B respectively, which is about the same as both the day-to-day and year-to-year variability, so more careful calibration will not significantly affect these results.

The MACHO light curves are publicly available via an interactive browser\footnote{http://www.macho.mcmaster.ca/Data/MachoData.html.}.  I derived standard {\it V} and {\it R} magnitudes from the instrumental magnitudes of MACHO's blue and red bands as \citet{b5} instructs.

The resultant light curves are plotted in Figs. 1 and 2.  For the MACHO data, some anomalous points are not within the plot range.  These figures show that the MACHO and OGLE times of observation overlap between MJD 50650 and MJD 51546.5, almost three seasons of data.  The relative strength of the periodicities during this time will be compared and discussed in terms of a physical interpretation.

For the following analysis I used the MACHO blue and red bands, not the standard bands.  The transformation required good data points for both bands simultaneously; by using the instrumental values I was able to retain the observations for which only one band had good data.

\begin{figure}
\scalebox{.55}{\includegraphics{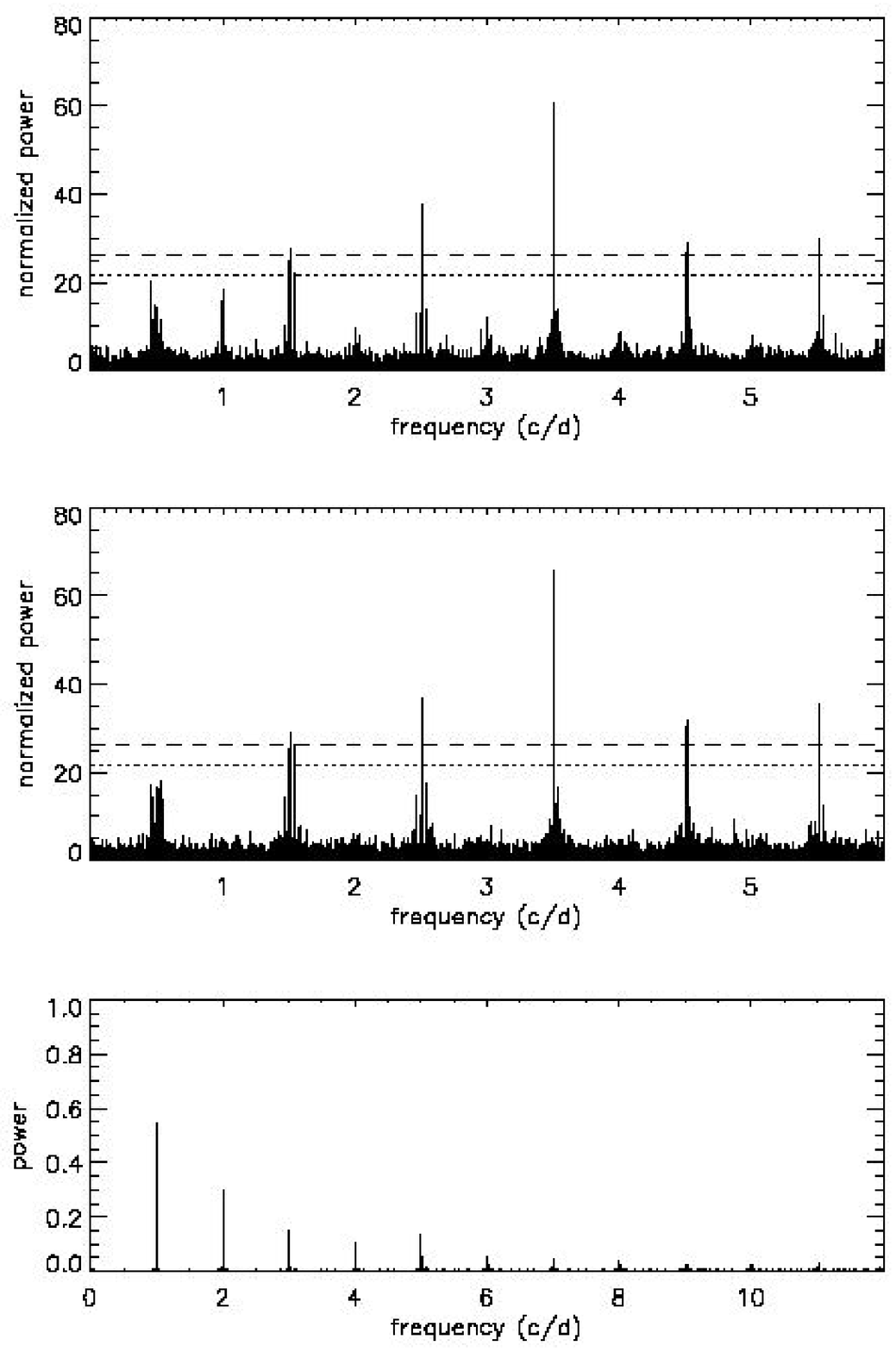}}
\caption{Star B's periodogram for MACHO blue (top) and red (middle).  The dotted line is 1\% false-alarm probability and the dashed line is 0.1\% false-alarm probability.  The spectral window (bottom) is very nearly the same for both.}
\scalebox{.55}{\includegraphics{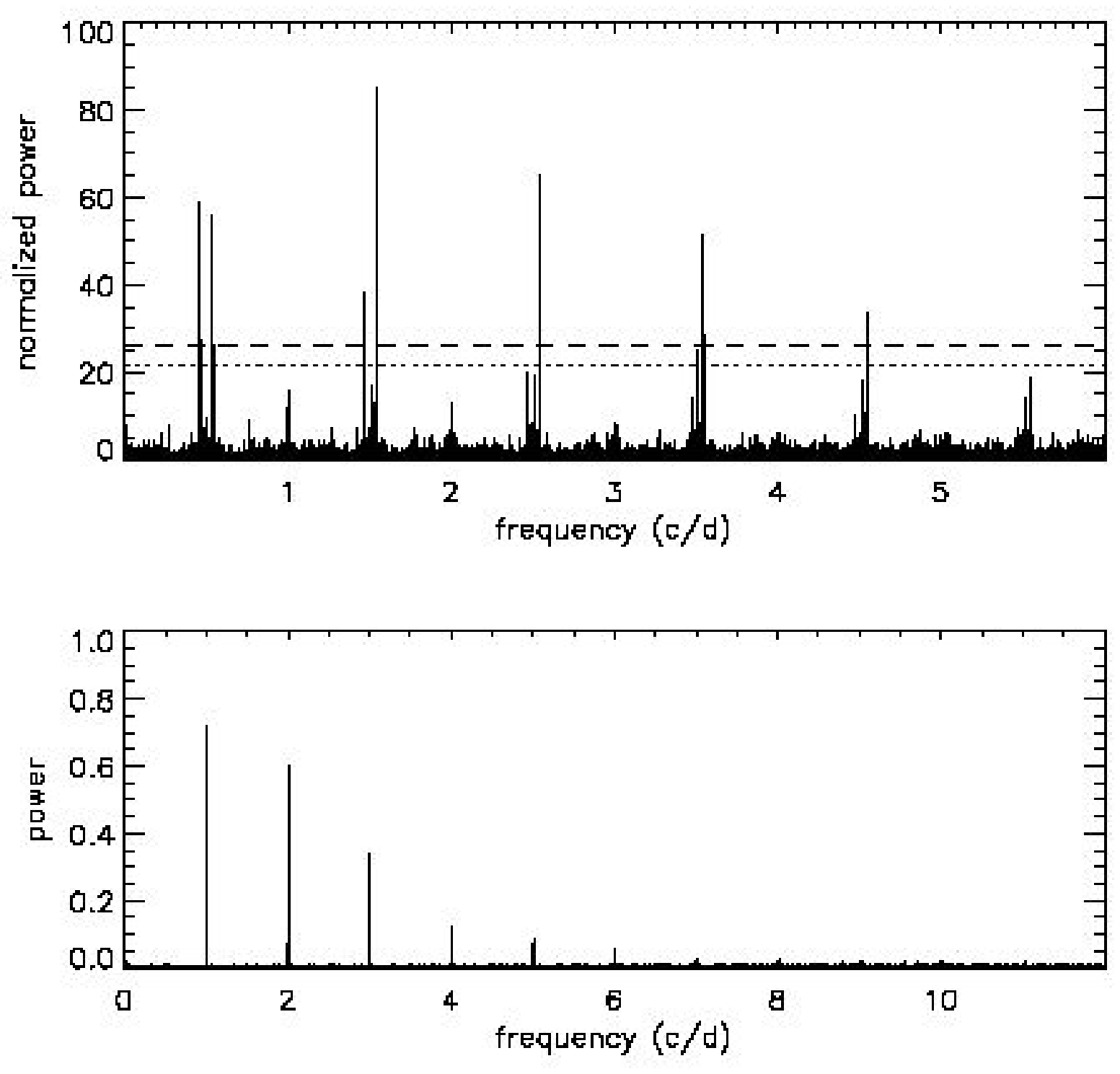}}
\caption{Same as above for OGLE {\it I} band.}
\end{figure}

\begin{figure}
\scalebox{.7}{\includegraphics{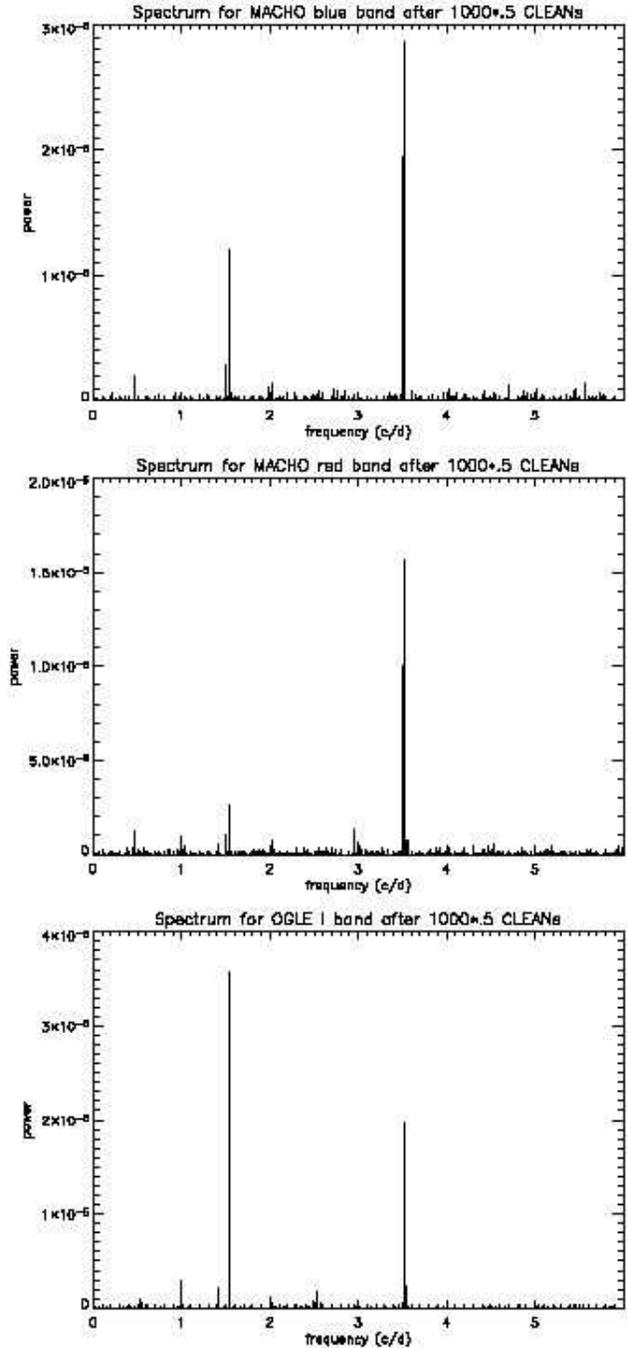}}
\caption{Top to bottom: MACHO blue and red, and OGLE {\it I} band CLEANed spectra.}
\end{figure}

\begin{figure}
\includegraphics{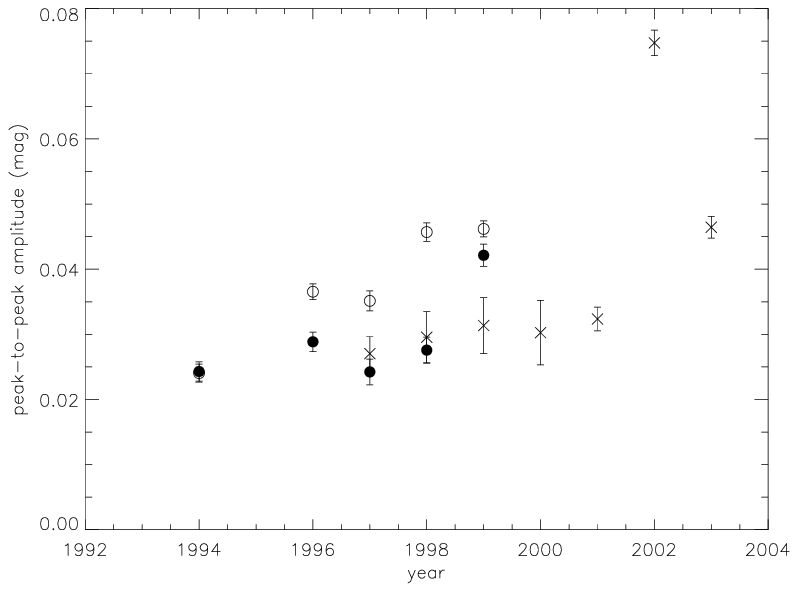}
\caption{The peak-to-peak amplitude of the periodic variation near 1.54 $c/d$, plotted by year for all data sets.  MACHO modulation in this region was not statistically significant during 1993 and 1995, so these years are omitted.  Errors were determined by finding the variance in the peak-to-peak amplitude of an ensemble of light curves created by varying each datum by a Gaussian of standard deviation equal to its error.  MACHO blue data are empty circles, MACHO red data are filled circles, and OGLE data are crosses.}
\includegraphics{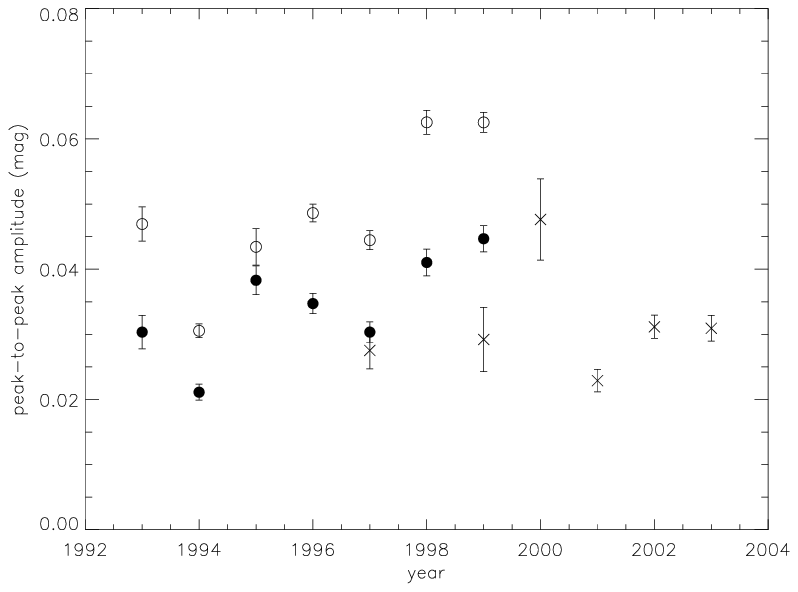}
\caption{Same as above, near 3.51 $c/d$ instead.  OGLE modulation in this region was not statistically significant during 1998, so this year was omitted.}
\end{figure}

\begin{figure}
\includegraphics{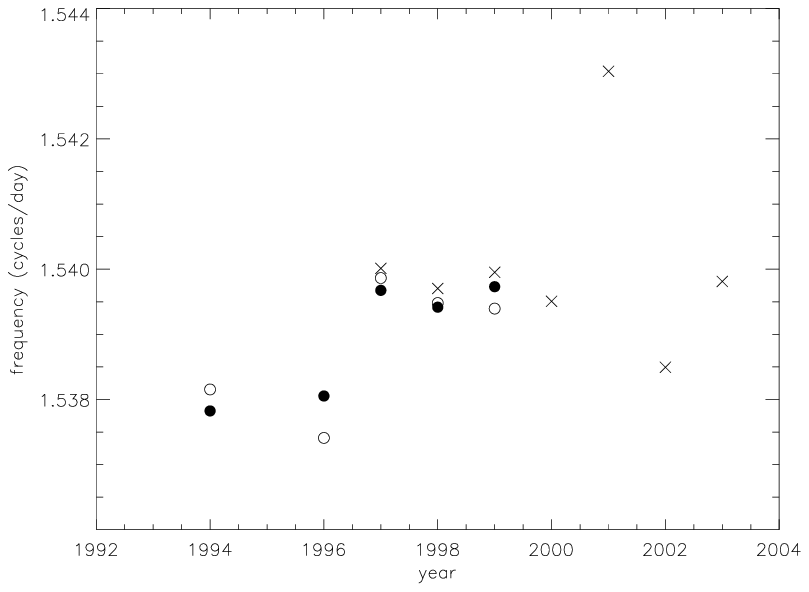}
\caption{$f_1$'s frequency by year, judged from the peak of the Lomb-Scargle periodogram.  MACHO modulation in this region was not statistically significant during 1993 and 1995, so these years are omitted.  MACHO blue data are empty circles, MACHO red data are filled circles, and OGLE data are crosses.  Typical FWHM is 0.004 $c/d$.}
\includegraphics{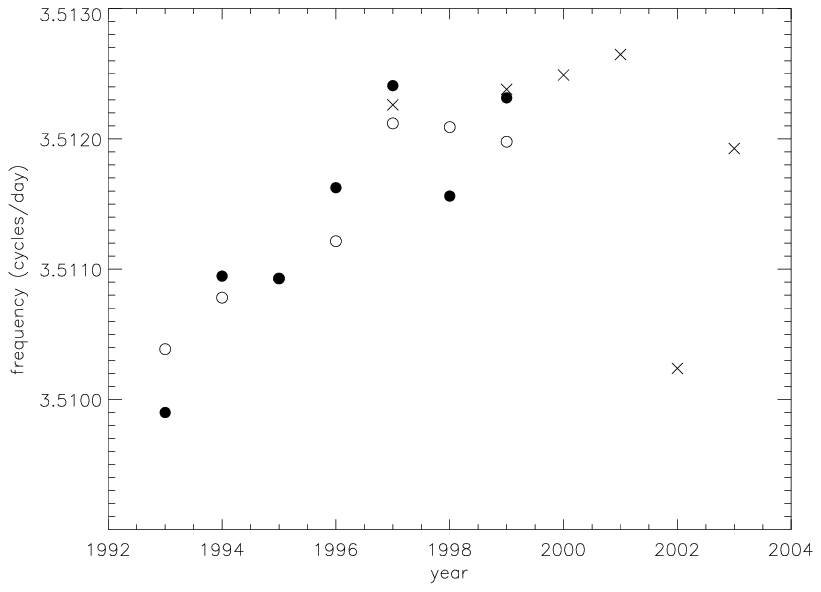}
\caption{Same as above, but for $f_2$.  OGLE modulation in this region was not statistically significant during 1998, so this year was omitted.}
\end{figure}

\section{Period Analysis}
As a first attempt to find periods, I made a Lomb periodogram using the modification of \citet{b7}.  

Fig. 1 shows changes in the brightness of star A with a time-scale of the length of the data, but whether it is periodic obviously cannot be determined.  Using the false-alarm probability from \citet{b19} to measure significance, none of the datasets yielded significant periods besides the power at a few hundred days, which can be seen in the light curve.  \cite{b2} proposed it as a Be star because it is within the error circle of an H-alpha emitting object \citep{b18}.  The light curve wanders on many time-scales, which is characteristic of Be variability, but it has a smaller amplitude than most Be stars.  It is possible that the H-alpha emission stopped after the observations of \cite{b18}, which were done in August of 1982, but before MACHO started taking data, so the photometric variations may represent the star and disk in a quiescent state.

For star B, the OGLE data showed a dominant period of 0.6494 $\pm$ 0.0001 days (1.5398 $\pm$ 0.0003 $cycles/day$, hereafter $f_1$).  Next I analyzed the MACHO data, which has peaks clustered around 0.2847 days (3.5119 $\pm$ 0.0010 $c/d$), hereafter $f_2$.  On closer inspection it became clear that both data sets had the other periodicity too, but it was lost in the aliases of the stronger period.

The spectra of evenly sampled data exhibit aliases at the significant frequency plus and minus multiples of the sampling frequency, and one can only reliably determine periods up to the Nyquist frequency, half the sampling frequency, keeping in mind that aliases from higher frequencies can appear.  For unevenly sampled data, aliasing does not render the periodicity ambiguous.  In that case, the true period can usually be gleaned from the peak of the spectrum since there is no solid sampling frequency.  Of course, for ground-based photometry, observations can only be taken at night, and this sampling introduces secondary peaks in the window function at multiples of one day.  If the telescope is in a location from which the star cannot be observed year-round, another small $year^{-1} =$ 0.002738 $c/d$ peak in the window function occurs.  Finally, the total length of the observations sets the width of a spectral feature, since two trial periods whose associated frequencies are closer than that will not go out of phase during the observation.  

Fig. 3 shows the periodogram and spectral window for the MACHO data of star B.  Plotted is the modulus squared of the spectral window; at negative frequencies reside the complex conjugate of the positive values.  Although bad points were removed only from the band in which they were bad, both bands share the same time sampling, so the window function is nearly identical.  As explained, the spectral window has its strongest aliases at multiples of 1 $c/d$, but they are weaker than the true value, so naively picking the peak as the true value works in this case.  Next to the $f_2 -$ 2 $c/d = $ 1.51 $c/d$ alias lies another significant peak at $f_1 = $ 1.54 $c/d$.  Its aliases can also be seen.  Fig. 4 is the periodogram and spectral window for the OGLE data for star B.  Here the $f_1$ peak is dominant with the $f_2$ peak hiding among its aliases.

\begin{figure*}
\scalebox{.8}{\includegraphics{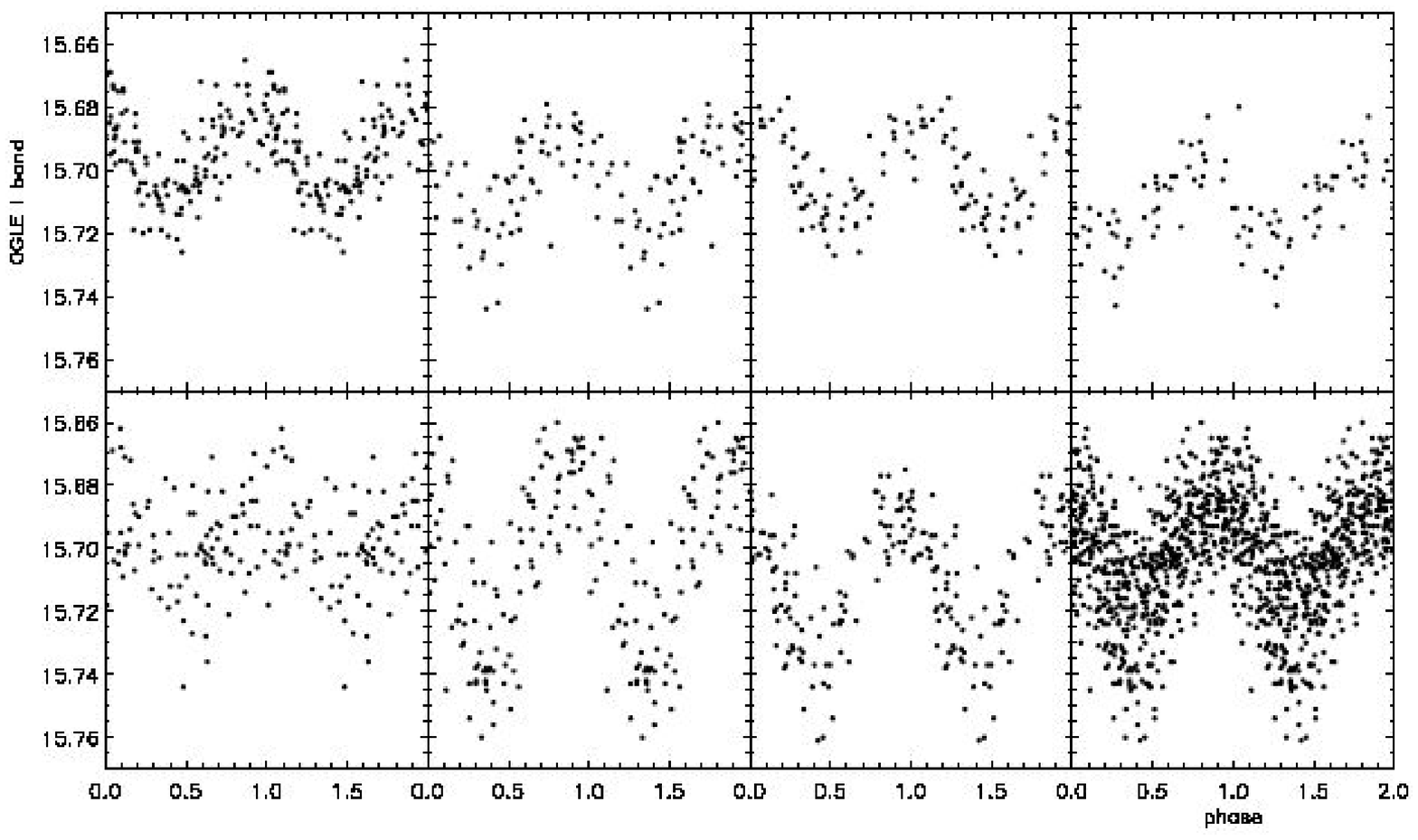}}
\caption{OGLE data for star B with power removed between 3.5023 and 3.5223 $c/d$, then folded at 1.5398 $c/d$.  Each plot is one year of data, starting in 1997.  The last plot is all seven years of data.  Errors are omitted for clarity.}
\scalebox{.8}{\includegraphics{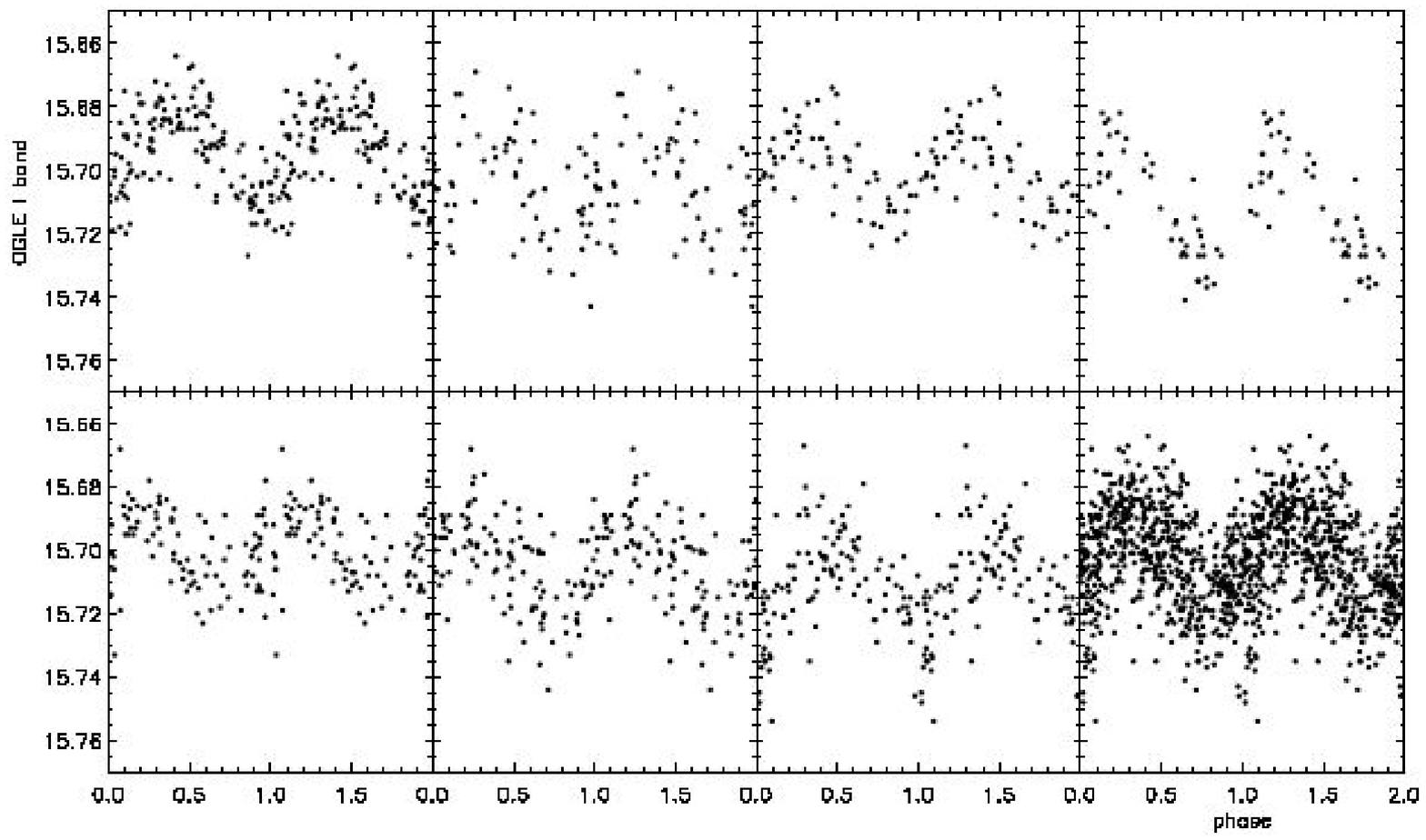}}
\caption{Same as above, but with power removed between 1.5298 and 1.5498 $c/d$, then folded at 3.5123 $c/d$, $f_2$'s peak during the OGLE years.}
\end{figure*}

The CLEAN algorithm was designed to extract multiple periodicities from unevenly sampled data despite its complicated window function \citep{b8}.  I ran CLEAN on all three datasets and discovered both frequencies present in all of them (see Fig. 5).  For each light curve the periodicities have different powers; this could be due to differences in both bandpass and epoch of observation.  In what follows I attempt to deconvolve this dual dependence.

I split the data into year-long chunks and analyzed them separately.  Due to its position at Las Campanas Observatory, Chile, the OGLE project is not able to monitor stars in the SMC all year round.  Most sets of observations began around June and went through February; for convenience I labeled such seasons with their starting year.  For instance, 2003 in Fig. 6-12 went until February, 2004.  I broke the MACHO data into similar chunks.  For each of these chunks, I recorded the peak frequency and amplitude in the vicinity (within 0.01 $c/d$) of the two modulations if it was significant at $\alpha = 0.01$ according to the false-alarm probability test of \citet{b19}.  Fig. 6 and 7 are the two modulations' amplitude plotted versus year.  For $f_2$, the bluer colours show considerably larger modulations, which is true for $f_1$ to a lesser degree.  See the discussion section for a possible interpretation.  The peak frequencies for each year are Fig. 8 and 9.

It is also instructive to view the folded light curves for each year.  If one plots the OGLE data folded at $f_2$, a faint pattern is discernible by eye.  If, however, the data is prewhitened with frequencies near $f_1$, the modulation is clear.  The CLEAN algorithm sequentially identifies and removes the strongest frequency, so as they were reported, if they fell within 0.01 $c/d$ of the dominant $f_1$ variation, I removed the best-fitting sinusoid of that frequency from the time series.  Fig. 10 and 11 show each year's OGLE data after this procedure: they are folded at each frequency after removing the other one.

\section{Phase Stability}

These yearly light curves, cleaned of the other period, also help probe the period stability of each variation.  I found that both periods were stable enough so that period stability can be described well as a change in phase with respect to a sine curve with a fixed frequency.

Fig. 12 has phase diagrams for each periodicity.  They were constructed by splitting up the yearly data into pieces of about 20 sequential observations.  The pieces never spanned OGLE's off-season between February and June.  Next, each piece was folded and a sinusoid was fit to it by least squares.  Plotted is the phase of that best-fitting sinusoid versus the mean time of observation.  A Monte-Carlo error estimate for that phase is also plotted: I varied each of the data points by a Gaussian of standard deviation equal to the quoted error and recalculated the phase, thus building up a list of 100 phases.  The standard deviation of that list is what I adopted as a measure of the error in phase.  Large error bars mean that offsetting the data had a large effect on the best fit, so in these cases it can be understood that the sinusoid fit the data only poorly from the start.  However, if the peak-to-peak amplitude of the fit was less than 10 mmag, I judged the phase unreliable and did not plot it. 

To give the plot continuity, I constrained the adjacent points in time for each band to have phase difference less than 0.5 cycles.  It should be noted that the phases plotted may be different by multiples of 1 $cycle$ from reality.  For instance, in the 1.5398 $c/d$ phase diagram in Fig. 12, year 1996 may not be properly matched to year 1995 due to sparse data and a quickly changing phase, and in the 3.5119 $c/d$ diagram, the first data points for each MACHO band may be plotted 1.0 cycle too high.

\begin{figure*}
\scalebox{.85}{\includegraphics{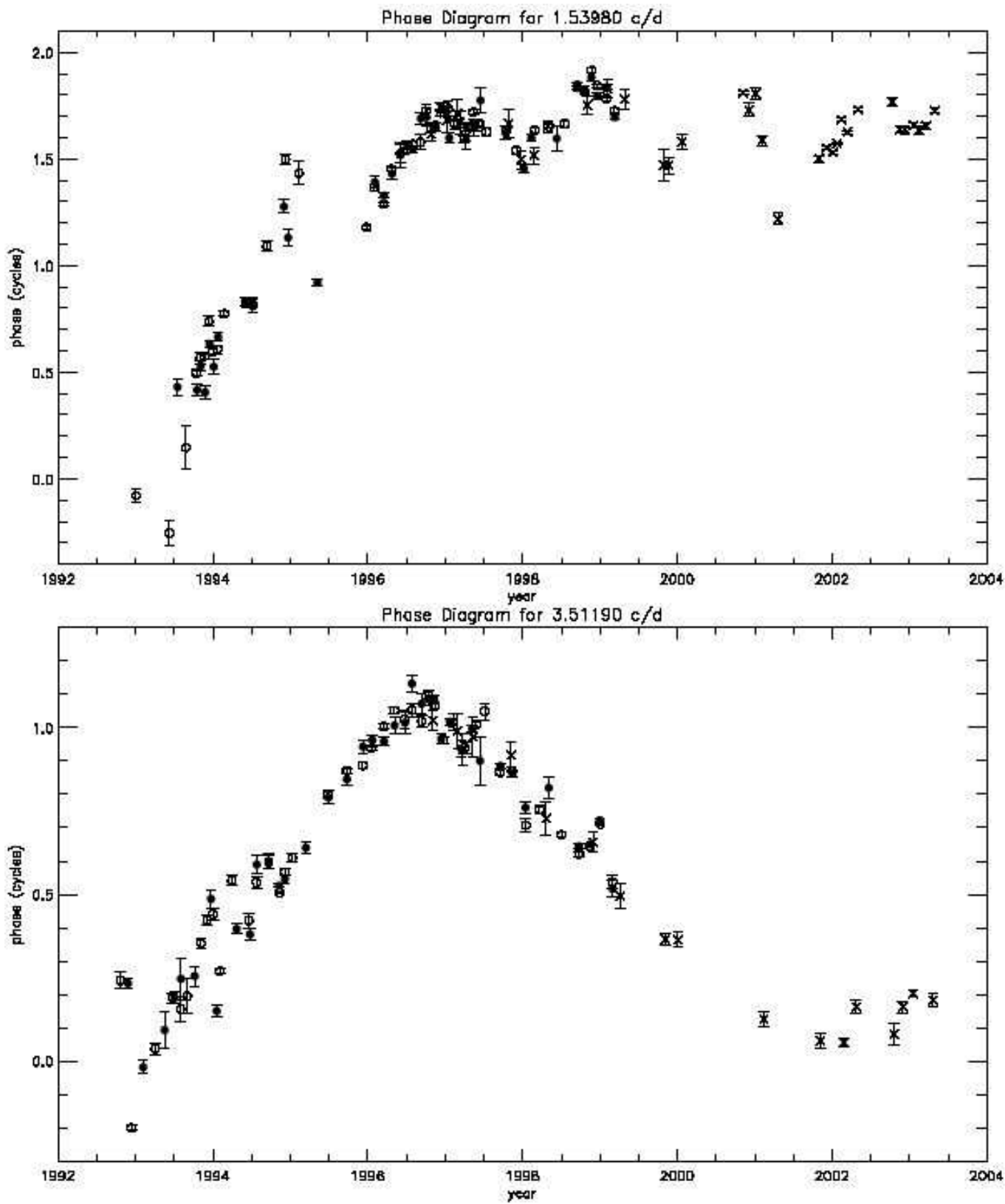}}
\caption{The data, after being cleaned of the other frequency, is split into chunks of about 20 observations and folded at each frequency.  The phase with a Monte-Carlo error estimate is plotted.  If the chunk has peak-to-peak amplitude less than 10 mmag, I judged the fit to be insignificant and did not plot a phase for that chunk.  Phase difference between adjacent points is forced under 0.5 cycles to impose continuity, but that decision is aesthetic: true phase may be an integer value off.  Open circles are MACHO blue, filled circles are MACHO red, and crosses are OGLE {\it I}-band.}
\end{figure*}

It is interesting to compare these phase diagrams with Figs. 10 and 11.  The bottom left box of Fig. 10 corresponds to the 2001 data.  For 2001 the peak of the periodogram is at 1.5430 $c/d$ instead of $f_1 = 1.5398 c/d$ (see Fig. 8).  This period decrease induced an advancing phase, so the folded plot does not look as crisp as other years.  If we consider the variation as simply a sinusoid with a discretely changing period, the change of about 0.003 $c/d$ would need about 200 days to build up to the 0.6 cycle advance shown for 2002 in the top plot of Fig. 12.  Then, in 2002, the peak of the periodogram is 1.5383 $c/d$, an apparently rather strong modulation and stable frequency, judging by the size of the error bars on Fig. 12 as well and the points' linearity, respectively.  The strength can be seen in the amplitude plot, Fig. 6.

A slight increase in frequency for both modulations occurred between the mid-90's and late-90's, as can be seen both in the frequency plots and the phase diagram.  It is unclear if a single physical mechanism modified both periods, but the possibility should be considered for a physical model.  It should also be noted that the phase matched across colors.  Some physical models may imply a detectable phase lag between band passes, but our analysis did not see any such lag.  

\section{Discussion}

\citet{b11} used Hipparcos photometry to discover that short-term periodic variability ($<$3.5 $days$) was present in about 40\% of Be stars between B0 and B3 and is less frequent in later types.  \citet{b2} calculated that both stars A and B have spectral type 09 V from the data in the catalog given by \citet{b10}, so the presence of a short-term period is not surprising. 

Neither of these periodicities are likely to be due to orbits around the star.  From the Corbet relation \citep{b9}, the expected orbital period is hundreds of days because it is correlated with the pulsar's long period of 701.6 $s$.  Also, if a solid body in Keplerian rotation were driving the variability, there would not have been glitches in phase stability.  For the same reason neither of the variations are likely to be caused by rotational modulation of a surface inhomogeneity, although such interpretation is the object of controversy \citep{b21}.

The periods should not be imputed to variations in the disk because the phase of the variation would not be so stable: disks in Be stars are transient.  It remains to be proven that there is a disk around star B; there was no coincident H-alpha emission in the catalog of \citet{b18}.  The only other (simple) explanation is stellar pulsations.  

Radial pulsations (i.e. $\beta$ Cephei stars) are a possibility.  The driving force for radial pulsation has been identified with an opacity ($\kappa$) mechanism of iron \citep{b16}, so as \citet{b17} pointed out, it should be a rarity the Magellanic Clouds.  In a recent OGLE-II study of the Large Magellanic Cloud \citep{b15}, three $\beta$ Cephei stars were found, and a study of the SMC for more was promised.  Periods from radial pulsations of the $\beta$ Cephei type in the Galaxy are generally less than 0.3 $d$ \citep{b14}, which fits for the shorter period of 0.285 $d$ but not for the longer period of 0.649 $d$.  However, \citet{b22} have found 20 doubly periodic B stars in the LMC with periods in the vicinity of $f_1$ and $f_2$, so this star may be the first SMC version of those stars.  The authors tentatively assigned those stars to both $\beta$ Cephei and Slowly Pulsating B-star types.  The following differences caution that assignment of star B to that group is non-trivial: this star has an earlier spectral type than any of those stars, it is in the SMC which has lower metallicity than the LMC, and it is part of a HMXB.

The blue band would be most sensitive to the temperature changes of a hot, radially pulsating star, and there is a marked increase in amplitude at higher wavelength for $f_2$.  During a cycle, the highest luminosity of such a star lags behind its highest temperature \citep{b20}.  This effect is unlikely to induce a detectable phase lag between our bandpasses, although this assertion should be quantified by future work.

Many recent studies describe the short term variability of Be stars (and hot stars in general) in terms of non-radial pulsations (NRP).  Such studies put an emphasis on time-resolved spectroscopy, looking for line-profile variability.  Photometric variations are thought to be less pronounced since the whole star and its disk contribute, possibly canceling out a locally strong variation.  \citet{b12} give line-profile variability for 27 early-type Be stars, and find that $l=|m|=2$ pulsations can explain nearly all those observations.  A few stars have been monitored extensively, and multiperiodicity has been detected in a few cases.  \citet{b13} found strong photometric periodicity of 1.56 $c/d$ and intermittent periodicity of various frequencies, including 3.23 $c/d$, in $\eta$ Cen.  The $\kappa$ mechanism has not been conclusively linked to NRP, so the low metallicity of the SMC is not necessarily a problem.  In line profile variations, features are seen traveling from blue to red \citep{b14}.  It remains to be determined if NRP would cause a detectable phase lag in the our colours.  The amplitude of variation is expected to depend on colour in NRP, but different parts of the star are relatively cold and hot simultaneously, so it is probably weaker than for radial pulsations.  NRP seems a likely source for the $f_1$ variability and a possible but unlikely source for the variation at $f_2$.

In conclusion, I suggest that the periodic variations of star B are due to stellar pulsations, but which ones cause them is left to future work.  Tentative evidence has been given that $f_1$ is a low-order non-radial pulsation and $f_2$ is a radial pulsation.  Spectroscopy and further analysis is needed before the cause of this multiperiodicity can be understood with rigor.  In particular, the star's metallicity is one important parameter to measure, and observations of line-profile variability would be very valuable to discern between radial and non-radial pulsation.

\section{Acknowledgments}

I thank Bohdan Paczy\'{n}ski for alerting me of these stars and advising me.  I also thank Andrzej Udalski and the OGLE team for providing their unpublished photometric data.

This paper utilizes public domain data obtained by the MACHO Project, jointly funded by the US Department of Energy through the University of California, Lawrence Livermore National Laboratory under contract No. W-7405-Eng-48, by the National Science Foundation through the Center for Particle Astrophysics of the University of California under cooperative agreement AST-8809616, and by the Mount Stromlo and Siding Spring Observatory, part of the Australian National University.

Soli Deo Gloria.

\label{lastpage}

\end{document}